\documentclass[onecolumn,showpacs]{revtex4}

\usepackage{graphicx}% Include figure files
\usepackage{dcolumn}% Align table columns on decimal point
\usepackage{bm}% bold math

\input epsf

\begin{document}

\title{\Large  VALIDITY OF THE GENERALIZED SECOND LAW OF THERMODYNAMICS OF THE UNIVERSE BOUNDED
BY THE EVENT HORIZON IN BRANE SCENARIO}

\author{\bf~Nairwita~Mazumder\footnote{nairwita15@gmail.com}, Subenoy~Chakraborty\footnote{schakraborty@math.jdvu.ac.in}.}

\affiliation{$^1$Department of Mathematics,~Jadavpur
University,~Kolkata-32, India.}

\date{\today}

\begin{abstract}
In this paper, we examine the validity of the generalized second
law of thermodynamics of the universe bounded by the event horizon
in brane-world gravity. Here we consider homogeneous and isotropic
model of the universe in one case where it is filled with perfect
fluid and in another case holographic dark energy model of the
universe has been considered.
\end{abstract}


\maketitle

\section{\normalsize\bf{Introduction}}

The discovery of Hawking radiation of black holes in the
semi-classical description, put black hole to behave as a black
body and there is emission of thermal radiation. The temperature
of the black hole (known as Hawking temperature) is proportional
to the surface gravity at the horizon while the entropy of the
black hole is related to the area of the horizon i.e. the above
thermodynamic quantities are related to the geometry of space-time
which is characterized by the Einstein field equations. So
naturally one may speculate some relationship between black hole
thermodynamics and Einstein equations. It is Jacobson [1] who
first showed that it is possible to derive Einstein field
equations from the first law of thermodynamics  for local Rindler
causal horizons. Subsequently, Padmanabhan [2] derived the first
law of thermodynamics on the horizon, starting from Einstein
equations for general static spherically symmetric space time.
Recently, to investigate the profound connection between gravity
and thermodynamics, Bamba et al [3] have examined the equivalence
of modified gravity equation to the clausius relation.\\

Subsequently , this nice equivalence between black hole
thermodynamics and the Einstein gravity raises the challenging
issue whether there is any relationship between the thermodynamics of
 space-time and the nature of gravity. In fact if we  assume the universe as a
 thermodynamical system bounded by the apparent horizon, its temperature
and entropy are given by $T_{A}=\frac {1} {2 \pi R_{A}}$,
$S_{A}=\frac{\pi {R_{A}}^2}{G}$ and the first law of
thermodynamics is equivalent with Friedmann equations  on the
apparent horizon. Also this equivalence still remains in higher
dimensional gravity theories i.e. f(R) gravity, Lovelock gravity
etc. In the context of brane world scenario, Ge [4] derived the
Friedmann-like equations for the RS-II model using the first
law of thermodynamics and Israel's junction conditions on the apparent horizon.\\

 So far, the thermodynamical analysis of the universe is mainly restricted to
the apparent horizon. In standard big bang cosmology event horizon
does not exist. However, the cosmological event horizon can be
distinguished from the apparent horizon in a general accelerating
universe dominated by dark energy with an equation of state
$\omega_{D}\neq -1$ . Further, using the usual definition of
temperature and entropy as in the apparent horizon  Wang et al [5]
have shown that both first and second law of thermodynamics break
down on the cosmological event horizon. They have argued that the
event horizon is a global feature of space-time while first law is
valid to nearby states of local thermodynamic equilibrium. Also it
is speculated that in the non-static universe, thermodynamical
quantities on the cosmological event horizon may not be as simple
as in the static space-time. Moreover, one may say the region
within the apparent horizon as the Bekenstein system because
Bekenstein's entropy mass bound $(S\leq 2 \pi R_{E})$ and
entropy-area bound ($S\leq \frac{A}{4}$) are satisfied in this
region. As Bekenstein bounds are universal and all gravitationally
stable special regions with weak self-gravity should satisfy
Bekenstein bounds so the corresponding
thermodynamical system is termed as a Bekenstein system.\\

In the context of recent observational evidences, the event
horizon exists due to accelerating phase of the universe. As the
universe bounded by the event horizon is not a Bekenstein system
so Bekenstein's entropy-area relation and also Hawking temperature
may not hold in the event horizon. Hence, from thermodynamical
point of view, it is interesting to determine the temperature and
entropy on the event horizon so that first law of thermodynamics
is always true there. As a first step to have some inside we
examine the validity of the generalized second law of
thermodynamics (GSLT) on the event horizon starting from the first
law of thermodynamics. The paper is organized as follows:
Effective  Friedmann equations in the brane scenario have been
formulated in section - II. The validity of GSLT has been examined
in sections III and IV for perfect fluid and holographic dark
energy respectively. The paper ends with  conclusions in
section V.\\

\section{\normalsize\bf{Randall-Sandrum -II Brane model and Effective Friedmann Equations:}}

Let us suppose that our n-dimensional visible universe is
described by an $(n-1)$-brane embedded in an $(n+1)$-dimensional
AdS space-time (called bulk). Due to string theory effects all
ordinary matters are confined in the brane and only gravity can
propagate across the brane. At each point on the brane, a
space-time unit normal (to the surface) $n^{A}$ can be defined as
$g^{AB}n_{A}n_{B}=1$ where $g_{AB}$ is the bulk metric. The
induced metric on the brane is given by

\begin{equation}
q_{\mu\nu}=g_{\mu\nu}-n_{\mu}n_{\nu}
\end{equation}

Here capital letters run over the $(n+1)$ bulk co-ordinates while
the greek indices corresponds to brane-coordinates. The brane can
be identified as the hypersurface $\chi=0$ where $\chi$ is a local
Gauss normal coordinate such that $n_{\mu}dx^{\mu}=d \chi$. If the
bulk is assumed to have only a cosmological constant
$\Lambda_{n+1}$, then the energy-momentum tensor in $(n+1)$
dimension has the form

\begin{equation}
T_{AB}=- \Lambda_{n+1} g_{AB} + S_{AB} \delta(x)
\end{equation}

with $$S_{\mu\nu}= - \lambda q_{\mu\nu} + \tau_{\mu\nu} .$$ Here
$\tau_{\mu\nu}$ is the energy-momentum tensor of the matter on
$(n-1)$ brane and $\lambda$ is the brane tension. The dirac delta
function assures the confinement of matter in the brane. Thus
considering brane as a singular hyper surface in the bulk we have
from the Israel's junction condition , the expression for the
extrinsic curvature

\begin{equation}
k_{\mu\nu}=- \frac{1}2{\kappa^{2}}_{n+1}
(S_{\mu\nu}-\frac{1}{(n-1)}q_{\mu\nu}S)
\end{equation}

where $Z_{2}$ symmetry (i.e. reflection symmetry) across the brane
has been imposed. Then using the Gauss equation and the bulk
Einstein equations the effective Einstein equations on the brane
has the form

\begin{equation}
G^{(n)}_{\mu\nu}=-\Lambda_{n}q_{\mu\nu}+8 \pi G_{n}T_{\mu
\nu}+\kappa^{4}_{(n+1)}\Pi_{\mu \nu}
\end{equation}

where $$ \Lambda_{n}= {\kappa^{2}}_{n+1}\left[ \frac{n-2}{n}
\Lambda_{n+1}+ \frac{(n-2)}{8(n-1)}
\kappa^{2}_{n+1}\lambda^{2}\right]$$ is the effective cosmological
constant on the brane,
$$G_{n}=\frac{(n-2)\lambda}{32\pi (n-1)}\kappa^{4}_{n+1},$$ is the
Newton's constant in $n-$dimensions and $$ \Pi_{\mu\nu}=-\frac{1}4
T_{\mu \alpha} T^{\alpha}_{\nu}+\frac{1}{4(n-1)}TT_{\mu
\nu}+\frac{1}8 g_{\mu \nu}T_{\alpha \beta}T^{\alpha
\beta}-\frac{1}{8(n-1)}T^{2}q_{\mu \nu}.$$\\

The above field equations on the brane may be considered as the
usual Einstein equations with an effective energy-momentum tensor,
given by the right hand side. One may note that the above
effective Einstein equations on the brane are not closed as the
bulk geometry can not be determined from these equations. Also the
standard Einstein equations can be recovered if
$\kappa_{n+1}\rightarrow0$, but $G_{n}$ is finite. Further, in the
low energy limit the quadratic correction term $\Pi_{\mu \nu}$
becomes insignificant and one gets back the usual Einstein
equations.\\

In the following sections, we mainly deal with the dynamical
aspects of the brane world cosmology so without any loss of
generality we assume the cosmological constant $\Lambda_{n}$
(independent of time) to be zero. As a result we have $$
\Lambda_{n+1}=-\frac{n}{8(n-1)}\kappa_{n+1}^{2}\lambda^{2}.$$ Let
us consider the $(n-1)-$brane to be $n-$dimensional FRW universe
embedded in a $(n+1)-$dimensional pure AdS bulk and the perfect
fluid is the matter contained in the brane. So the metric on the
brane can be written as

\begin{equation}
ds^{2}=q_{\mu \nu}dx^{\mu}dx^{\nu}=
-dt^{2}+\frac{a^2(t)}{1-kr^2}dr^2 + a^2(t) r^2 d\Omega_{n-2}^{2}
\end{equation}
$$~~~~~~=~h_{ab}dx^{a}dx^{b}~+~{\tilde{r}}^{2}d\Omega_{n-2}^{2}$$
where $\tilde{r}=ar$ is the area radius, $x^0=t$, $x^1=r$ and
~$h_{ab}=~diag(-1,\frac{a^{2}}{(1-kr^{2})})$ is the metric of the
$2$-space.\\

Using the form of the energy-momentum tensor

\begin{equation}
T_{\mu \nu}=(\rho+p)u_{\mu}u_{nu}+p~q_{\mu \nu}
\end{equation}

with $u_{\mu}u^{\mu}=-1$, the quadratic correction term $\Pi_{\mu
\nu}$ has the expression

\begin{equation}
\Pi_{\mu \nu}=\frac{(n-2)}{8(n-1)}\rho
~[~2(\rho+p)u_{\mu}u_{\nu}+(\rho+2p)~q_{\mu \nu}]
\end{equation}

Thus the explicit effective Einstein equations on the brane are
given by

\begin{equation}
(n-1)(n-2)\left(H^2+\frac{k}{a^2}\right) = 16 \pi G_{n} \left(
\rho+ \frac{(n-2)}{(n-1)}\frac{\kappa^{4}_{n+1}}{64\pi G_n} \rho^2
\right)
\end{equation}

\begin{equation}
-(n-2)\left(\dot{H}-\frac{k}{a^2}\right) = 8 \pi G_{n} \left(
\rho+p+ \frac{(n-2)}{(n-1)}\frac{\kappa^{4}_{n+1}}{32\pi G_n} \rho
(\rho+p) \right)
\end{equation}

The energy conservation equation takes the form
\begin{equation}
\dot{\rho}+(n-1)H(\rho+p)=0
\end{equation}

\section{\normalsize\bf{Study of GSLT on the event horizon:}}

The FRW brane and the effective Einstein equations on it are
presented in the previous section. The apparent horizon is
characterized by the relation

\begin{equation}
h^{ab}{\tilde{r}}_{,~a}{\tilde{r}}_{,~b}=0
\end{equation}

and we have the radius of the apparent horizon

\begin{equation}
R_{A}=\frac{1}{\surd(H^{2}+\frac{k}a^{2})}
\end{equation}

where $H=\frac{\dot{a}}a$ is the usual Hubble parameter and
$k=0,\pm1$ for flat ,closed and open FRW model.\\

Assuming the existence of the event horizon (which exists for an
accelerated expanding universe in Einstein gravity) the radius of
the cosmological event horizon is given by

\begin{equation}
R_{E}=a\int^{\infty}_{t}\frac{dt}{a}=a\int^{\infty}_{a}\frac{da}{Ha^{2}}
\end{equation}

Thus the apparent horizon, event horizon and the Hubble horizon
$(R_{H}=\frac{1}H)$ are connected as described in Ref.[6] for flat
open and closed model for the universe.\\

Now we project the effective energy-momentum tensor on the brane
in the normal direction to the $(n-1)$space and define

\begin{equation}
W \equiv -\frac{1}2 {\tilde{T}}^{ab}h_{ab}
\end{equation}

and
\begin{equation}
\psi_{a} \equiv {\tilde{T}}^{a}_{b} \partial_{b}\tilde{r} +
W\partial_{a} \tilde{r}
\end{equation}

where ${\tilde{T}}^{ab}$ is the projected energy-momentum tensor ;
$W$ and $\psi_{a}$ are called the work density and the energy
supply vector [7] respectively. The work density $W$ at the
horizon may be viewed as the work done by the change of the
horizon while the energy-supply at the horizon is total energy
flow through the horizon .The area $(A)$ and volume $(V)$ of the
$(n-1)$brane of radius $\tilde{r}$ having explicit form

\begin{equation}
A=(n-1)\Omega_{n-1}\tilde{r}^{n-2}~~ and~~
V=\Omega_{n-1}\tilde{r}^{n-1}
\end{equation}

with $\Omega_{n-1}=\frac{\Pi^{\frac{(n-1)}2}}{\Gamma(\frac{n-1}2 +
1)}$, volume of an $(n-1)$ D unit ball.\\

Let us now consider the heat flow $\delta Q$ through the event
horizon for an infinitesimal time $\delta t$, assuming the volume
of the brane to be unchanged (i.e. $dV=0$). As this heat flow
$\delta Q$ is equivalent to the amount of energy crossing the
event horizon in time $\delta t$ so we have from clausius relation
[8]

\begin{equation}
T_{E}dS_{E}=\delta Q=-dE=-A\psi
\end{equation}

where ~$S_{E}$~ and $T_{E}~$ are the entropy and temperature on
the event horizon.\\

For the present brane model the explicit expressions of work
density and energy supply are given by

\begin{equation}
W=-\frac{1}{2}\left(-\rho+p+\frac{n-2}{4(n-1)}\frac{\kappa_{n+1}^4}{\kappa_n^2}\rho
p\right)dt
\end{equation}
and

\begin{equation}
\psi=-\frac{\kappa_n^2 R_{E}H}{8 \pi
G_n}(\rho+p)\left(1+\frac{n-2}{4(n-1)}\frac{\kappa_{n+1}^4}{\kappa_n^2}\rho\right)dt
\end{equation}

and we have

\begin{equation}
dS_{E}=\frac{(n-1)\Omega_{n-1}R_E^{n-1}H\kappa_n^2} {T_E 8 \pi
G_n}(\rho_{total}+p_{total})dt
\end{equation}

From Gibb's equation [9]

\begin{equation}
T_{E}dS_{I}=dE_{I}+pdV
\end{equation}

and we have

\begin{equation}
dS_{I}=\frac{-(n-1)\Omega_{n-1}R_E^{n-2}(\rho+p)}{T_E}dt
\end{equation}

where $S_I$ is the entropy of the matter distribution bounded by
the event horizon and $E_I$ is the energy of the inside matter
with expression $$E_I=\Omega_{n-1}R_E^{n-1} \rho~.$$ It is to be
noted that for thermodynamical equilibrium the temperature of the
matter distribution is chosen as that of the boundary (the event
horizon). Further, in deriving the expression for $dS_I$ from
Gibb's equation, we have used the variation of the event horizon

\begin{equation}
\frac{dR_E}{dt}=\left(R_E-\frac{1}H\right)H
\end{equation}

Thus using the energy conservation relation the time variation of
the total entropy is given by

\begin{equation}
\frac{d}{dt}(S_E+S_I)=(n-1)\Omega_{n-1}(\rho+p)
\frac{R_E^{n-2}H}{T_E}\left[R_E\left(1+\frac{\rho}{\lambda}\right)-\frac{1}H\right]
\end{equation}

\section{\normalsize\bf{GSLT on the event horizon of (n-1)Brane filled with Holographic Dark Energy(HDE):}}

Let us consider the flat FRW model of the universe filled with
holographic dark energy [10].We have the modified Friedmann
equation in $(n-1)-$brane is

\begin{equation}
(n-1)(n-2)\left(H^2+\frac{k}{a^2}\right) = 16 \pi G_{n} \left(
\rho_D+ \frac{(n-2)}{(n-1)}\frac{\kappa^{4}_{n+1}}{64\pi G_n}
\rho_D^2 \right)
\end{equation}

\begin{equation}
-(n-2)\left(\dot{H}-\frac{k}{a^2}\right) = 8 \pi G_{n} \left(
\rho_D+p_D+ \frac{(n-2)}{(n-1)}\frac{\kappa^{4}_{n+1}}{32\pi G_n}
\rho_D (\rho_D+p_D) \right)
\end{equation}

where $\rho_D$ and $p_D$ are the energy density and thermodynamic
pressure corresponding to HDE. The energy conservation equation
takes the form
\begin{equation}
\dot{\rho_D}+(n-1)H(\rho_D+p_D)=0
\end{equation}

where the effective holographic dark energy density [11,12] is
given as

\begin{equation}
\rho_D=c^2{(\sqrt{ \pi }M_n)}^{n-3}M_n A_n^{-1}
\frac{n-2}{8\Gamma(\frac{n-1}2)}R_E^{-2}
\end{equation}

where $M_n=G_n^{-\frac{1}{n-2}}~,~G_n~$is the $n-$ dimensional
gravitational constant ; $A_n=\frac{\Pi
^{\frac{n-1}2}}{\left(\frac{n-1}2\right)!}$ and $c^2$ is a
numerical factor.\\

Now the amount of energy crossing the event horizon during time
$dt$ is given by

\begin{equation}
-dE=(n-1)\Omega_{n-1}R_E^{n-1}H(\rho_t+p_t)
\end{equation}

where $\rho_t$ and $p_t $ is the effective energy density and
pressure in brane world scenario with

\begin{equation}
\rho_t+p_t=
(\rho_D+p_D)\left(1+\frac{n-2}{4(n-1)}\frac{\kappa_{n+1}^4}{\kappa_n^2}\rho_D\right)
\end{equation}
So assuming the validity of the first law of thermodynamics we get

\begin{equation}
dS_{E}=\frac{(n-1)\Omega_{n-1}R_E^{n-1}H\kappa_n^2} {T_E 8 \pi
G_n}(\rho_{D}+p_{D})\left(1+\frac{n-2}{4(n-1)}\frac{\kappa_{n+1}^4}{\kappa_n^2}\rho_D\right)dt
\end{equation}

Now taking logarithm on both sides of equation (28) and
differentiating we get

\begin{equation}
dR_E=\frac{3}2(1+\omega_D)R_E Hdt
\end{equation}

where the equation of state of HDE is $p_D=\omega_D \rho_D$ ,
$\omega_D$ is not necessary a constant.\\

To obtain the variation of the entropy of the fluid inside the
event horizon of the brane we use as before the Gibb's equation
(21) with $$E_I=\Omega_{n-1}R_E^{n-1} \rho_D~~
and~~V=\Omega_{n-1}R_E^{n-1}$$ and we get

\begin{equation}
\frac{dS_{I}}{dt}=\frac{(n-1)\Omega_{n-1}{R_{E}}^{n-1}H(\rho_{D}+p_{D})}{T_{E}}
\left[\frac{n-1}2(1+\omega_{D})-1\right]dt
\end{equation}

Therefore combining (31) and (33) the time variation of total
entropy is given by

\begin{equation}
\frac{d}{dt}(S_{I}+S_{E})=\frac{(n-1)\Omega_{n-1}{R_{E}}^{n-1}H(\rho_{D}+p_{D})}{T_{E}}
\left[\frac{\rho_D}{\lambda}+\frac{n-1}2(1+\omega_D)\right]
\end{equation}

where we have used
$$\frac{\kappa_{n+1}^4}{\kappa_n^2}=\frac{4(n-1)}{(n-2) \lambda}$$
which is the relation between the brane tension and the
gravitational coupling constants in bulk and brane.\\

\section{\normalsize\bf{Conclusions:}}

In the last two sections we have derived expressions for total
entropy change with respect to time for perfect fluid and
holographic dark energy model respectively. he conclusions for
the validity of GSLT are as follows:\\

For flat $(k=0)$ or open $(k=-1)$ FRW model $R_E>\frac{1}H$, so if
the matter satisfies weak energy condition then GSLT is always
valid. However, for closed $(k=+1)$ FRW model if $R_E>\frac{1}H$
then as before validity of GSLT depends on the weak energy
condition. But if for closed model $R_E<\frac{1}H$ then also weak
energy condition is sufficient for the GSLT to be satisfied
provided the expression within the square bracket in equation (24)
is positive definite , otherwise GSLT will be valid if the matter
do not obey the weak energy condition (i.e. of phantom nature).
For validity of GSLT in phantom thermodynamics
one may refer to reference [14].\\

Moreover, if $\lambda$ is sufficiently small so that the
expression within the square bracket is positive for all FRW model
(flat,open,closed) then matter should be non-exotic is the only
criteria for the validity of the GSLT.\\

For holographic dark energy model if $\omega_D>-1$ then GSLT is
always satisfied i.e. if the holographic dark energy is not exotic
then GSLT will be valid. However, for large $\lambda$ the first
term within the square bracket in equation (34) may be
insignificant and then GSLT is satisfied without any
restriction on the nature of the holographic dark energy.\\

It is to be noted that for validity of the GSLT we have used the
first law of thermodynamics with the assumption that the volume of
the brane does not change for an infinitesimal time, but we have
no need to use any explicit form of temperature and entropy on the
event horizon. However, for equilibrium thermodynamics, we have
chosen the temperature of the matter
inside the event horizon is the same as the horizon temperature.\\

If brane tension $\lambda$ becomes very large then results for
both the two cases reduces to those for Einstein
Gravity [see Ref. [5] and [13] ].\\

In the literature there are works related to negative brane
tension (see ref [15] and there in). Brane model with negative
brane tension is commonly known as bouncing brane [16]. Also for
$\lambda <0$ a phantom universe that begins with an accelerated
phase can be evolved into a decelerated phase. In the present
context if $\lambda<0$ then validity of GSLT is as follows: For
perfect fluid and holographic model there is an upper bound for
energy density in quintessence era while for phantom model the
energy density has a lower bound (the magnitude of the brane
tension) for perfect fluid and GSLT is always valid for HDE
model.\\

The criteria for the validity of GSLT in brane scenario differ
insignificantly from that in Einstein gravity. In particular, for
small brane tension $\lambda$, the condition for validity of GSLT
for closed FRW model with perfect fluid as the matter is different
from that for Einstein gravity. Also for holographic dark energy
model there is no restriction for the validity of GSLT for
Einstein gravity while in brane scenario the
matter should obey the weak energy condition.\\

The matter on the brane is assumed to be such that
produces accelerated expansion so that event horizon exists.\\

We shall now make a comparative study of the restrictions for the
validity of GSLT that we have obtain in this paper for Brane
scenario with those for general relativity and in other gravity
theory. First of  all for perfect fluid the general restriction is
that "GSLT is satisfied for open and flat model provided weak
energy condition is satisfied while for closed model, in addition
some restrictions among the horizon radii is also needed". This
result is true for Einstein gravity [5], f(R) gravity and modified
gravity with logarithmic correction in entropy-area relation [17].
However, in the present work in Brane scenario the restriction for
GSLT is quite distinct and it dependents on the brane tension to a
great extended particularly for closed model. For small $\lambda$
the expression within the square bracket (in eq. (24)) become
positive so that GSLT can not be satisfied in phantom era. (In
scalar tensor theory it is very complicated.[18])

In the second case when universe is filled with holographic dark
energy then GSLT is always satisfied both in quintessence and in
phantom era without any restriction [13]. This result is true in
Einstein gravity [13], F(R) gravity, modified gravity with
logarithmic correction in entropy-area relation [17] and also in
DGP brane model [19]. But in the present brane model we need
restrictions for validity of GSLT. The term containing brane
tension within square bracket (in equation (34)) plays a crucial
role for the validity of GSLT. Therefore, we may conclude that the
restrictions for the validity of GSLT are quite distinct in brane
scenario compared to those in other gravity theories.

For further work , it will be interesting to examine the validity
of GSLT for interacting two fluid system and make a comparative
study with the corresponding result in general relativity [20] and
DGP brane model [21]. Also the first law of thermodynamics on the
event horizon will be examined to have any idea about the entropy
and temperature in the event horizon.\\\\

{\bf Acknowledgement:}\\

One of the author (SC) is thankful to CSIR, Govt. of India for
providing a project ("Cosmological Studies in Brane World
Scenario"). NM is thankful to CSIR for providing JRF in the
project.\\

{\bf References:}\\
\\
$[1]$ T.Jacobson , \it {Phys. Rev Lett.} {\bf 75} 1260 (1995) \\\\
$[2]$ T. Padmanabhan , {\it Class. Quant. Grav. } {\bf 19} 5387 (2002).\\\\
$[3]$ Kazuharu Bamba, Chao-Qiang Geng, Shin'ichi Nojiri and Sergei
D. Odintsov, {\it Euro Phys. Lett.} {\bf 89} (2010) 50003\\\\
$[4]$ Xian-Hui Ge , {\it Phys. Lett. B} {\bf 651} (2007)
49.\\\\
$[5]$ B. Wang, Y. Gong, E. Abdalla , \it{Phys. Rev. D} {\bf 74}
083520 (2006).\\\\
$[6]$ N. Mazumder and S. Chakraborty , {\it Class. Quant. Gravity}
{\bf 26} 195016 (2009).\\\\
$[7]$ R.G. Cai and L.M. Cao , {\it Phys. Lett. B} {\bf 785} (2007)
135.\\\\
$[8]$ R.G. Cai and L.M. Cao , {\it Phys. Rev. D} {\bf 75} 064008 (2007).\\\\
$[9]$ G. Izquierdo and D. Pavon , {\it Phys. Lett. B} {\bf 633}
420 (2006).\\\\
$[10]$ M. Li , {\it Phys. Lett. B} {\bf 603} 01 (2004);\\\\
$[11]$ E.N. Saridakis , {\it Phys. Lett. B } {\bf 660} (2008) 138.\\\\
$[12]$ K. Ke and M. Li , {\it Phys. Lett. B} {\bf 606} 173 (2005);\\\\
$[13]$ N. Mazumder and S. Chakraborty , {\it Accepted in Gen.Rel.Grav.} doi:10.1007/s10714-009-0881-z.\\\\
$[14]$ Shin'ichi Nojiri, Sergei D. Odintsov, {\bf
arXiv: hep-th/0505215}.\\\\
$[15]$ Samuel Lepe, Francisco Peña, Joel Saavedra {\it Physics
Letters B} {\bf 662}217
(2008)\\\\
$[16]$Y. Shtanov, V. Sahni, {\it Phys. Lett. B} {\bf 557} (2003)
1, gr-qc/0208047\\\\
$[17]$ N. Mazumder and S. Chakraborty,  arXiv:1005.5215
[gr-qc]\\\\
$[18]$ N. Mazumder and S. Chakraborty, arXiv:1005.5217 [gr-qc]\\\\
$[19]$ J. Dutta, S. Chakraborty and M. Ansari,  arXiv:1005.5321
[gr-qc]; J. Dutta and S. Chakraborty, {\it General Relativity and
Gravitation},
{\bf DOI:10.1007/s10714-010-0957-9}.\\\\
$[20]$ N. Mazumder and S. Chakraborty, arXiv:1005.5589 [gr-qc].\\\\
$[21]$ J. Dutta and S. Chakraborty,  arXiv:1006.2210 [gr-qc].\\\\

\end{document}